\documentclass[aps,preprint,amsmath,amssymb]{revtex4}
\usepackage{graphicx,color}

\def\be{\begin{eqnarray}}
\def\ed{\end{eqnarray}}
\def\ga{\gamma}
\def\non{\nonumber}
\def\la{\langle}
\def\ra{\rangle}

\begin{document}
\title{\Large \bf Axigluon on like-sign charge asymmetry ${\cal A}^b_{s\ell}$, FCNCs and CP asymmetries in $B$ decays    }
\date{\today}
\author{ \bf  Chuan-Hung Chen$^{1,2}$\footnote{Email:
phychen@mail.ncku.edu.tw} and Gaber
Faisel$^{3,4}$\footnote{Email:gfaisel@cc.ncu.edu.tw}  }
 \affiliation{$^{1}$Department of Physics,
National Cheng-Kung University, Tainan, 701
Taiwan\\
 $^{2}$National Center for Theoretical Sciences, Hsinchu 300, Taiwan \\
 $^{3}$ Egyptian Center for Theoretical Physics, Modern University
for Information and Technology, Cairo, Egypt \\
 $^{4}$Physics Department, Faculty of Education, Thamar
University, Thamar ,Yemen }

\begin{abstract}
A non-universal axigluon in generalized chiral color models leads to flavor changing neutral currents (FCNCs) at tree level. We analyze phenomenologically the new contributions to  $B_{q}$ (q=d, s) mixing and the related CP asymmetries (CPAs) that are generated by axigluon exchange. We find that although $\Delta m_{B_q}$ can give a strict constraint on the parameters of $b\to q$ transition,  the precise  measurement of $\sin2\beta_{J/\Psi K^0}$ can further exclude the parameter space of $b\to d$ transition. The axigluon-mediated effects can enhance the like-sign dimuon charge asymmetry ${\cal A}^{b}_{s\ell}$  by one order of magnitude larger than the standard model prediction. Accordingly,   large CPA $\sin2\beta^{J/\Psi \phi}_s$ and CPA difference $\sin2\beta_{J\Psi K^0}-\sin2\beta_{\phi K^0}$  are achieved.

\end{abstract}

\maketitle

\section{Introduction}

In the standard model (SM), with three families of quarks, the
unique CP violating phase of  the Cabibbo-Kobayashi-Maskawa (CKM)
matrix  can  explain some of the observed CP violating phenomena
in $K$ and $B$ systems. However, the failure of the KM phase in
explaining the matter-antimatter asymmetry and some recent
measurements of CP violating observations in $B$ meson mixings and
decays motivates the search for new source of CP violation (CPV).
Therefore, it is an important issue to explore and to find  new CP
violating effects in various systems, such as cosmos, Large Hadron
Collider (LHC), Tevatron, $B$ factories etc.

Recently, several hints for the existence of new CP violating
sources are revealed in experiments. The first hint is observed in
the CP asymmetries (CPAs) of $ B \to \pi K$ decays where by naive
SM estimation, one expects that $\bar B_d\to \pi^+ K^-$ and $
B^-\to \pi^0 K^-$ decays have similar CPAs. However, it is
surprising that the world average difference between the two CPAs
contradicts the expectation as the experimental result is
\cite{TheHeavyFlavorAveragingGroup:2010qj}
 \be
\Delta A_{CP}= A_{CP}(\pi^+ K^{-}) - A_{CP}(\pi^0 K^-)= -(14.8_{-1.4}^{ +1.3})\%\,,
 \ed
 whereas the SM prediction is  $\Delta A_{CP}(SM) =  0.025 \pm
0.015$ \cite{Beneke:2003zv}. The  large deviation from the SM
prediction indicates a puzzle in the asymmetries and it is
introduced in the literature as $B\to \pi K$ puzzle \cite{pikpuz}.
The second hint is observed in  the time-dependent CPA of $B_s$
system, where CDF and D{\O} have shown an unexpected large CP
phase in the mixing-induced CPA for $B_s\to J/\Psi \phi$ and the
two possible solutions are given by \cite{TheHeavyFlavorAveragingGroup:2010qj}
 \be
2\beta^{J/\Psi \phi}_{s}=2\beta_s + 2\phi^{\rm NP}_{s}= -0.75^{+
0.32}_{-0.21}\ {\texttt{or}}\  -2.38^{+0.25}_{-0.34}
 \ed
at $90\%$ confidence level (CL). Here, $\beta_s\approx -0.019$
\cite{Chen:2008ug} is the SM CP violating phase  and
$\phi^{\rm NP}_s$ is the CP violating phase of  new physics. The
significant deviation from the SM prediction could be speculated
by the contributions of new physics.

The third hint is observed in the like-sign charge asymmetry which
is defined as \cite{Abazov:2010hv}
 \be
{\cal A}^{b}_{s\ell} &=& \frac{N^{++}_b - N^{--}_{b}}{N^{++}_{b}+N^{--}_{b}}\,, \label{eq:Absl_exp}
 \ed
where $N^{++(--)}_b$ denotes the number of events that $b$- and
$\bar b$-hadron semileptonically decay into two positive
(negative) muons.  Recently,  D{\O} Collaboration has announced
the measurement on ${\cal A}^{b}_{s\ell}$ in the dimuon events
\cite{Abazov:2010hv} with
 \be
 {\cal A}^{b}_{s\ell} = \left(-9.57 \pm 2.51({\rm stat}) \pm 1.46 ({\rm syst}) \right)\times 10^{-3}\,.
 \ed
The SM prediction is ${\cal A}^{b}_{s\ell}=(-2.3^{+0.5}_{-0.6})\times 10^{-4}$
\cite{Abazov:2010hv,Lenz:2006hd}. If the semileptonic b-hadron
decays do not involve CP violating phase, then the charge
asymmetry is directly related to the mixing-induced CPAs in $B_d$-
and $B_s$-meson oscillations. Although the errors of the data are
still large, however the $3.2$ standard deviations from the SM prediction can be attributed to new CP violating phases in $b\to d$ and $b\to s$ transitions \cite{Randall:1998te,Dighe:2010nj,Dobrescu:2010rh,Choudhury:2010ya}.

In order to explore the new physics and to avoid the uncontrollable QCD uncertainties, we will concentrate our study on the mixing parameter $\Delta m_{B_q}$,  the charge asymmetry ${\cal A}^{b}_{s\ell}$ and  the  time-dependent CPA in $B_q$ oscillation, where QCD effects can be controlled well by Lattice QCD.

In the literature, many extensions of the SM such as chiral color
models \cite{chiral,nonuni,Sehgal:1987wi,Doncheski:1997yj,Giordani:2003ib,Choudhury:2007ux}, $Z'$ models \cite{LL_PRD45,BZ1,BZ2} etc have been proposed. The flavor non-universal axigluon in the generalized chiral color models \cite{AKR,Frampton:2009rk} has been studied for solving the anomalous forward-backward asymmetry (FBA) in the top-quark pair production at the Tevatron \cite{D0_PRL100, CDF_PRL101}. Although other models such as Z', diquarks models \cite{Arhrib:2009hu} etc may have  significant contributions to the FBA, however,  large gauge couplings and flavor changing effects should be introduced in which chiral color model does not need.  Inspired by the effects of the axigluon on the top-quark FBA, we study the axigluon-mediated phenomena in $B$-meson system.

A flavor universal axigluon  has flavor-conserving effects only. For non-universal axigluon which has different couplings to different quarks, flavor changing neutral currents (FCNCs) can be generated at tree level. This is achieved after transforming the weak eigenstates of the quarks into their physical eigenstates. As a consequence, many phenomena will be affected by these FCNC effects. In this paper, we analyze in detail the non-universal axigluon contributions to the time-dependent CPAs in $B_q$ oscillation after taking into account the constraint from the mixing parameter $\Delta m_{B_q}$.

This paper is organized as follows. In Sec.~\ref{sec:formalism},
we formulate the interactions of $b\to q$ transitions  which are
induced by flavor non-universal axigluon exchange. Accordingly, we
derive the corresponding effective Hamiltonian for $\Delta B=1,2$
processes. Furthermore, we discuss the contributions of the
axigluon to the charge asymmetry ${\cal A}^{b}_{s\ell}$ and the
time-dependent CPAs for $B_d\to J/\psi K^0$, $B_d\to \phi
K^0$, $B_s\to J/\Psi \phi$ decays. The detailed numerical analysis is
presented in Sec.~\ref{sec:NA}. We give the conclusion in
Sec.~\ref{sec:conclusion}.

\section{Formalism}\label{sec:formalism}

In order to study the contributions of the non-universal axigluon
to the FCNC processes, we start by writing the interactions of the
massive color-octet gauge boson with quarks as
  \be
{\cal L}_{A} &=& g_V \bar q' \ga_\mu T^b q' G^{b\mu}_{A} + g_A \bar q' \ga_\mu \ga_5
{\bf Z} T^b q' G^{b\mu}_{A}\,, 
  \ed
where we have suppressed the flavor and color indices, $g_{V,A}$ are the gauge couplings of the new gauge group $SU(3)_A\times SU(3)_B$, $T^b$ are the Gell-Mann matrices which are normalized by $Tr(T^b T^c)= \delta^{ac}/2$ and $\bf Z$ is $3\times 3$ diagonalized matrix with diag(Z)=(1, 1, $\zeta$). Here $\zeta =\tilde g_A/g_A$ where $\tilde g_A$ denotes the gauge coupling of
the third-generation quark and its value depends on a specific model, e.g. $\zeta=-1$ in Ref.~\cite{Frampton:2009rk}.  For simplicity, we assume that the new exotic quarks which are required for anomaly free are very heavy and their effects are negligible. Hence, we still focus on three flavors for each up and down type quarks. Following the scenario in Refs.~\cite{AKR,Frampton:2009rk} for solving the large top-quark FBA, we assume that the axigluon couplings to the third generation are different from their couplings to the first two generations. The left- and right-handed quarks are $SU(2)$ doublet and singlet respectively. Thus, after spontaneous symmetry breaking, the interacting and physical eigenstates can be related by unitary matrices as $q_\chi= V^Q_\chi q'$ with $\chi$ being the chiralities $L$ and $R$ and $Q$ being  up or down type quarks.
Since $\bf Z$ is not a unit matrix,  the FCNCs are arisen from the
axial-vector currents and the corresponding Lagrangian is given by
 \be
{\cal L}_{FCNC}&=& g_A \bar q \ga_\mu  ( V^Q_R {\bf Z} V^{Q\dagger}_{R} P_R - V^Q_L {\bf Z}
V^{Q\dagger}_{L} P_L ) T^b q G^{b\mu}_{A} \label{eq:FCNC}
 \ed
with $P_{L(R)}=(1\mp \ga_5)/2$. Since $V^Q_\chi$ are unknown
matrices, the FCNCs are associated with left and right-handed
currents generally. Nevertheless, if $V^Q_R = V^Q_L$, from
Eq.~(\ref{eq:FCNC}) we see that  the FCNCs are only associated
with axial-vector currents. In terms of the flavor indices, the
matrix $V^q_\chi {\bf Z} V^{q\dagger}_{\chi}$ can be decomposed as
 \be
 \left(V^Q_\chi {\bf Z} V^{Q\dagger}_{\chi} \right)_{ij} &=& \delta_{ij} + \left(V^Q_\chi ({\bf Z- 1} )
  V^{Q\dagger}_{\chi} \right)_{ij} = \delta_{ij} + (\zeta-1) (V^Q_{\chi})_{i 3} (V^{Q*}_{ \chi})_{ j3}\,.
 \ed
Therefore, the  Lagrangian of $b\to q$ transition can be written
as
 \be
{\cal L}_{b \to q}&=&  g_A \bar q \ga_\mu  ( F^{QR}_{qb} P_R - F^{QL}_{qb} P_L ) T^b b G^{b\mu}_{A} \label{eq:bq}
 \ed
with $F^{Q\chi}_{qb} = (\zeta-1)(V^Q_{\chi})_{i 3} (V^{Q*}_{
\chi})_{ 33}$ where $i=(1,2,3)$ denotes the family order of the
same type $Q$ quark. Based on Eq.~(\ref{eq:bq}), we study the
impacts of non-universal axigluon exchange on $\Delta B=2$
processes and the time-dependent CPAs in $B_q$ system.

By Eq.~(\ref{eq:bq}), the effective Hamiltonian for $\Delta B=2$
transitions  which is generated by the tree-level axigluon
mediation can be written as
 \be
{\cal H}^A_{\Delta B=2} &=& \frac{g^2_A}{4m^2_V} \left[ -\frac{1}{N_C}\left( \bar q \ga_\mu (F^{DR}_{qb}
P_R+F^{DL}_{qb} P_L)b \right)^2 \right. \non \\
&+& \left. \bar q_\alpha \ga_\mu \left(F^{DR}_{qb} P_R+F^{DL}_{qb} P_L \right)b_\beta \bar q_\beta \ga^\mu
\left(F^{DR}_{qb} P_R+F^{DL}_{qb} P_L \right)b_\alpha \right]\,,
 \ed
where $N_C$ denotes the number of colors and we have used the identity
 \be
T^b_{ij} T^b_{k\ell} = -\frac{1}{2N_C} \delta_{ij} \delta_{k\ell}
+ \frac{1}{2} \delta_{i\ell} \delta_{jk}\,. \label{eq:su3}
 \ed
In order to calculate the $B_q-\bar B_q$ mixing, we write the relevant
hadronic matrix elements to be
\begin{eqnarray}
\langle  B_q| \bar q \gamma_{\mu} P_{L(R)} b \bar q \gamma_{\mu} P_{L(R)} b | \bar B_q \rangle
= \frac{1}{3} m_{B_q} f^2_{B_q} \hat B_q \,, \non \\
\langle   B_q| \bar q \gamma_{\mu} P_{R} b \bar q \gamma_{\mu} P_{L} b | \bar B_q \rangle
= -\frac{5}{12} m_{B_q} f^2_{B_q} \hat B^{RL}_{1q} \,, \non \\
\langle B_q| \bar q_\alpha \ga_\mu P_{L} b_\beta \bar q_\beta  \ga^\mu P_{R} b_\alpha | \bar B_q \rangle
= - \frac{7}{12}  m_{B_q} f^2_{B_q}  \hat B^{RL}_{2q}\,.
\end{eqnarray}
 To estimate the new physics effects, we
employ the vacuum insertion method to calculate the above matrix
elements, i.e. $\hat B_q \sim \hat B^{RL}_{1q} \sim \hat
B^{RL}_{2q}\sim 1$ \cite{Gabbiani:1996hi,Badin:2007bv}.  Additionally, in the
heavy quark limit, we  take $m_b \sim m_{B_q}$. As a result, the
transition matrix element for $B_q-\bar B_q$ oscillation mediated
by axigluon exchange
 becomes
 \be
 M^{A,q}_{12} &=& \langle B_q | {\cal H}^A_{\Delta B=2}| \bar B_q \rangle = \frac{g^2_A}{18 m^2_V} m_{B_q} f^2_{B_q} U^D_{qb}\,, \non \\
 U^D_{qb} &=& (F^{DR}_{qb})^2 +(F^{DL}_{qb})^2 + 4 F^{DR}_{qb} F^{DL}_{qb}\,. \label{eq:Uqb}
 \ed
For reducing the number of free parameters, we will take the
approximation $V^{Q}_R \approx V^{Q}_L =V^D$ in our analysis, i.e.
$F^{DR}_{qb} \approx F^{DL}_{qb} = F^D_{qb}$, then $U^{D}_{qb}= 6
(F^{D}_{qb})^2$. We note that the approximation $V^{Q}_R \approx
V^{Q}_L$ can be realized in hermitian Yukawa matrices
\cite{Chen:2001cv}.

By combining the contributions of SM and axigluon, the transition matrix element for $\Delta B =2$ can be formulated as
 \be
M^{B_q}_{12}
%
&=&|M^{\rm SM,q}_{12}| R^q_A  e^{i2(\beta_q + \phi^{\rm NP}_{q})}\,,
 \ed
where the new parameters are defined by
 \be
R^q_A &=& \left(1+(r^q_A)^2 + 2r^q_A \cos2(\beta^{\rm NP}_{q}-\beta_q)\right)^{1/2}\,,\non \\
2\beta^{\rm NP}_q &=& {\rm arg}(M^{A,q}_{12})\,, \non \\
r^q_A &=& \frac{|M^{A,q}_{12}|}{|M^{\rm SM,q}_{12}|}\,, \non \\
\tan2\phi^{\rm NP}_{q} &=& \frac{r^q_A \sin2(\beta^{\rm NP}_q -\beta_q)}{1+r^q_A \cos2(\beta^{\rm NP}_{q} -\beta_q)}\,, \label{eq:mBq}
 \ed
and  $M^{SM,q}_{12}$ is given by \cite{BBL}
 \be
M^{SM,q}_{12}=\frac{G^{2}_{F} m^2_{W}}{12 \pi^2}
\eta_{B} m_{B_q}f^{2}_{B_q}\hat{B}_q ( V^*_{tq}V_{tb})^2 S_{0}(x_t)
 \ed
with $S_{0}(x_t)=0.784 x_t^{0.76}$, $x_{t}=(m_t/m_W)^2$ and
$\eta_{B}\approx 0.55$ is the QCD correction to $S_0(x_t)$
Hence, the mass difference between heavy and light $B_q$ is
$\Delta m_{B_q} = 2|M^{B_q}_{12}|=\Delta m^{\rm SM}_{B_q} R^q_A$.
After obtaining $M^{B_q}_{12}$,  the time-dependent CPA through
inclusive semileptonic decays can be defined as \cite{Nakamura:2010zzi}
 \be
a^q_{s\ell}&=& \frac{\Gamma(\bar B_q(t) \to \ell^+ X)- \Gamma( B_q(t) \to \ell^- X)}{\Gamma(\bar B_q(t) \to \ell^+ X)
+\Gamma( B_q(t) \to \ell^- X)}\,,\non \\
&=& \frac{1-|q/p|^4}{1+|q/p|^4}
 \ed
with
 \be
 \left(\frac{q}{p}\right)^2 &=& \frac{ M^{B_q^*}_{12}-i\Gamma^{B_q^*}_{12}/2}{ M^{B_q}_{12}-i\Gamma^{B_q}_{12}/2}\,,
 \ed
where $\Gamma^{B_q}_{12}$ denotes the absorptive part of $B_q
\leftrightarrow \bar B_q$ transition. Due to $\Gamma^{B_q}_{12}\ll
M^{B_q}_{12}$, the wrong-sign charge asymmetry can be simplified
as
 \be
a^q_{s\ell} &=& Im \left( \frac{\Gamma^{B_q}_{12}}{M^{B_q}_{12}}\right)\approx \frac{\Delta
\Gamma^{\rm SM}_{B_q}}{\Delta m_{B_q}}\sin(2\beta_q +2\phi^{\rm NP}_{q} - \theta^\Gamma_q)\,. \label{eq:aqsl}
 \ed
Here, $\theta^\Gamma_q$ stands for the phase of $\Gamma^{B_q}_{12}$. Since the absorptive part is  dominated by the SM contribution, we will assume that $\Gamma^{B_q}_{12}= \Gamma^{q,SM}_{12}$ in our numerical analysis.  A detailed discussions about new physics effects on $\Gamma^{B_q}_{12}$ can be found in Refs.~\cite{Dighe:2010nj,Choudhury:2010ya}. Since $a^{q}_{s\ell}$ is associated with the CP phases directly, a non-zero charge asymmetry will be an indication of CP violation. Accordingly, the like-sign charge asymmetry defined in Eq.~(\ref{eq:Absl_exp}) can be written as \cite{Abazov:2010hv, Grossman:2006ce}
 \be
 {\cal A}^b_{s\ell} &=& \frac{\Gamma(b\bar b\to \ell^+ \ell^+ X) - \Gamma(b\bar b\to
 \ell^- \ell^- X)}{\Gamma(b\bar b\to \ell^+ \ell^+ X) + \Gamma(b\bar b\to \ell^- \ell^- X)}\,,\non\\
 &=& \frac{f_d Z_d a^d_{s\ell} + f_s Z_s a^s_{s\ell}}{f_d Z_d + f_s Z_s}\,, \label{eq:Absl_1}
 \ed
where $f_q$ is the production fraction of $B_q$ and
 \be
 Z_q &=& \frac{1}{1-y^2_q} -\frac{1}{1-x^2_q}\,, \non \\
 y_q &=& \frac{\Delta\Gamma_{B_q}}{2\Gamma_{B_q}}\,, \ \ \
 x_q = \frac{\Delta m_{B_q}}{\Gamma_{B_q}}\,.
 \ed
Using $f_d=0.323(37)$, $f_s=0.118(15)$, $x_d =0.774(37)$, $y_d\sim 0$, $x_s=26.2 (5)$ and $y_s=0.046(27)$, the asymmetry can be rewritten as
 \be
{\cal A}^b_{s\ell} = c_d a^d_{s\ell} + c_s a^s_{s\ell} \label{eq:Absl_2}
 \ed
with $c_d=0.506(43)$ and $c_s=0.494(43)$ \cite{Abazov:2010hv}.

Another important time dependent CPA can be defined  by \cite{Nakamura:2010zzi}
 \be
A_{f_{CP}}(t)&=& \frac{\Gamma(\bar B_q(t) \to f_{CP})- \Gamma( B_q(t) \to f_{CP})}{\Gamma(\bar B_q(t) \to f_{CP})
+\Gamma( B_q(t) \to f_{CP})}\,,\non \\
&=& S_{f_{CP}} \sin\Delta m_{B_q} t - C_{f_{CP}} \cos\Delta m_{B_q} t\,, \non\\
S_{f_{CP}} &=& \frac{2Im\lambda_{f_{CP}} }{1+|\lambda_{f_{CP}}|^2}\,,\ \ \ C_{f_{CP}}
= \frac{1-|\lambda_{f_{CP}}|^2}{1+|\lambda_{f_{CP}}|^2} \label{eq:Sf}
 \ed
with
 \be
 \lambda_{f_{CP}} &=&-\left(\frac{M^{B_q^*}_{12}}{M^{B_q}_{12}}\right)^{1/2}
  \frac{A(\bar B\to f_{CP})}{A(B\to f_{CP})} =  -e^{-2i(\beta_q +
  \phi^{\rm NP}_{q}) }\frac{\bar A_{f_{CP}}}{A_{f_{CP}}}\,, \label{eq:lambdaf}
 \ed
where $f_{CP}$ denotes the final CP eigenstate, $S_{f_{CP}}$ and
$C_{f_{CP}}$ are the so-called mixing-induced and direct CPAs,
$A_{f_{CP}}$ and $\bar A_{f_{CP}}$ are the amplitudes of $B$ and
$\bar B$ mesons decaying to $f_{CP}$ and $\bar
A_{f_{CP}}/A_{f_{CP}}=-\eta_{f_{CP}} A_{f_{CP}}(\theta_W\to
-\theta_W)/A_{f_{CP}} (\theta_W)$ with $\eta_{f_{CP}}$ and
$\theta_W$ are the CP eigenvalue of $f_{CP}$ and the weak CP phase
respectively. Clearly, besides $\Delta B=2$ effects, the
mixing-induced CPA is also related to the $\Delta B=1$ process. In
this paper, we will concentrate on $f_{CP}=J/\Psi K_S$ and $\phi
K_S$ for $q=d$ and on $f_{CP}=J/\Psi \phi$ for $q=s$.

To calculate the decay amplitude of $B(\bar B)\to f_{CP}$, we need
to discuss the interactions of $\Delta B=1$ processes. With the
approximation $V^{Q}_R \approx V^{Q}_L$, the effective Hamiltonian
of $b\to q q' q'$ can be expressed as
 \be
{\cal H}_{b\to qq'q'} &=& \frac{g_A}{m^2_V} F^D_{qb} \bar q \ga_\mu \ga_5 T^b b
\sum_{q'=u,d,s,c} \bar q' \ga^\mu \left(g_+ P_R + g_{-} P_L \right) T^b q'
 \ed
with $g_{\pm}= g_{V} \pm  g_{A}$. Using Eq.~(\ref{eq:su3}), we can rewrite the last equation  as
 \be
{\cal H}^A_{b\to qq'q'} &=& \frac{G_F }{\sqrt{2}}V^{*}_{tq}V_{tb} \left[
C'_{q3} O^q_3 + C'_{q4} O^q_4 +C^R_{q3} O^{qR}_3 + C^R_{q4} O^{qR}_4 \non \right.\\
&+& \left.C'_{q5} O^q_5+C'_{q6} O^q_6 +C^L_{q5} O^{qL}_{5} + C^{qL} O^{qL}_6\right] \label{eq:bqint}
 \ed
in which the new Wilson coefficients are expressed by
 \be
C'_{q3} &=&\frac{1}{8N_C} \frac{\sqrt{2}F^D_{qb}}{G_F V^{*}_{tq}V_{tb}} \frac{g_A g_{-}}{m^2_V}\,, \ \ \  C'_{q4}=-N_C C'_{q3}\,, \non\\
C^L_{q5}&=&-C'_{q3}\,, \ \ \ C^L_{q6}= -N_C C^L_{q5}\,, \non \\
C'_{q5} &=&\frac{1}{8N_C} \frac{\sqrt{2}F^D_{qb}}{G_F V^{*}_{tq}V_{tb}} \frac{g_A g_{+}}{m^2_V}
\,, \ \ \ C'_{q6} = - N_C C'_{q5} \,, \non \\
C^R_{q3} &=& - C'_{q5}\,,\ \ \ C^R_{q4} = -N_C C^R_{q3} \label{eq:coeff}
 \ed
and the associated operators are
 \be
O^q_3 &=&  (\bar q b)_{V-A} \sum_{q'} (\bar q' q')_{V-A}\,, \ \ \ O^q_4 =  (\bar q_\alpha b_\beta)_{V-A} \sum_{q'}
(\bar q'_\beta q'_\alpha)_{V-A}\,, \non \\
O^q_5 &=&  (\bar q b)_{V-A} \sum_{q'} (\bar q' q')_{V+A}\,, \ \ \ O^q_6 =  (\bar q_\alpha b_\beta)_{V-A} \sum_{q'}
(\bar q'_\beta q'_\alpha)_{V+A}\,, \non \\
O^{qR}_3 &=& (\bar q b)_{V+A} \sum_{q'} (\bar q' q')_{V+A}\,, \ \ \ O^{qR}_4 =  (\bar q_\alpha b_\beta)_{V+A} \sum_{q'} (\bar q'_\beta q'_\alpha)_{V+A}\,,\non\\
O^{qL}_5 &=& (\bar q b)_{V+A} \sum_{q'} (\bar q' q')_{V-A}\,, \ \ \ O^{qR}_6 =  (\bar q_\alpha b_\beta)_{V+A} \sum_{q'} (\bar q'_\beta q'_\alpha)_{V-A}
 \ed
with $(\bar f' f)_{V\pm A} =\bar f' \ga_\mu (1\pm \ga_5) f$. Besides the new free parameters that are introduced earlier, the  non-leptonic $B$ decays suffer from large uncertain QCD effects such as $\la f_{CP} | {\cal
H}_{b\to qq'q'}| B\ra$. For estimating the new physics effects, we
employ the naive factorization approach (NFA). Under the NFA, we
find that the related effective Wilson coefficients for $\bar B_d
\to J/\Psi \bar K^0$ and $\bar B_s\to J/\Psi \phi$ are
 \be
  C'_{s3} + \frac{C'_{s4}}{N_C} + C^L_{s5} + \frac{C^R_{s6}}{N_C}+C'_{s5} + \frac{C'_{s6}}{N_C} + C^R_{s3} + \frac{C^R_{s4}}{N_C}\,. \label{eq:zeroCs}
 \ed
With the results in Eq.~(\ref{eq:coeff}), we clearly see that the
influence of axigluon-mediated effects on $J/\Psi (\bar K^0, \phi)$
modes vanishes. In our analysis we neglect the nonfactorizable
 contributions as they are subleading and difficult to
estimate. Now, only $B\to \phi K$ can display the
axigluon-mediated effects. Using NFA and the interactions of
Eq.~(\ref{eq:bqint}), the total decay amplitude of $B\to \phi K$
is written as
 \be
\bar A_{\phi \bar K^0}&=& \la \phi \bar K^0 | {\cal H}_{b\to ss\bar s}|\bar B^0 \ra\,, \non\\
&=& \frac{G_F}{\sqrt{2}} V^*_{ts} V_{tb} (a^{\rm SM}+a'_{s4} + a^R_{s4}) \la \phi|\bar s \ga_\mu s|0\ra \la \bar K^0| \bar s \ga^\mu b|\bar B\ra
 \ed
where ${\cal H}_{b\to ss\bar s}$ is the sum of the SM and axigluon
effective Hamiltonian and  $a^{\rm SM}=a_3+ a_4 + a_5$ with
 \be
 a_3&=&C_3 +\frac{C_4}{N_C}\,, \ \ \ a_4 = C_4 + \frac{C_3}{N_C}\,, \ \ \  a_5 = C_5 + \frac{C_6}{N_C}\,,\non \\
&& a'_{s4}=C'_{s4} + C'_{s3}/N_C\,, \ \ \ a^{R}_{s4}= C^R_{s4} + C^R_{s3}/N_C\,. \non
 \ed
Here, $C_{3-6}$ are the effective Wilson coefficients from the
gluon penguin in the SM \cite{BBL}. We note that the
electroweak penguin contributions are very small
and thus we neglect them. Using $V_{ts}=-|V_{ts}|e^{-i\beta_s}$
\cite{Nakamura:2010zzi}, we can write
 \be
\frac{\bar A_{\phi \bar K^0}}{A_{\phi K^0} }= -e^{2i\beta_s} \frac{a^{SM} + a^{R}_{s4}}{a^{SM} + a^{R^*}_{s4}}=-e^{2i(\beta_s
 + \theta^{\rm NP}_{s})}  \ed
with
 \be
\tan\theta^{\rm NP}_{s} &=& \frac{|a^R_{s4}| \sin(\beta^{\rm NP}_{s}-\beta_s) }{a^{SM} + |a^{R}_{s4}| \cos(\beta^{\rm NP}_{s} -\beta_s)}\,. \non
 \ed
By Eqs.~(\ref{eq:Sf}) and (\ref{eq:lambdaf}), the mixing-induced CPA via $B_d\to \phi K^0$ decay is obtained as
 \be
 S_{\phi K^0} &\equiv& \sin2\beta_{\phi K^0} = \sin2(\beta_d + \phi^{\rm NP}_{d}-\beta_s
 -\theta^{\rm NP}_{s})
 \,,
 \label{eq:S_phik}
 \ed
while the CPAs through $B_{d,s}\to J/\Psi (K_S, \phi)$ decays are given by
 \be
 S_{J/\Psi K^0}&\equiv& \sin2\beta_{J/\Psi K^0} = \sin2(\beta_d + \phi^{\rm NP}_{d})\,, \non \\
 S_{J/\Psi \phi} &\equiv& \sin2\beta^{J/\Psi \phi}_{s} = \sin2(\beta_s + \phi^{\rm NP}_{s})\,. \label{eq:Sjpsi_phi}
 \ed
Although the measurement of $\sin2\beta_{J/\Psi K^0}$ has
approached to the precision level, however, it might be difficult
to tell if there exists new physics by measuring
$\sin2\beta_{J/\Psi K^0}$ only. Nevertheless, one can investigate
 a new asymmetry defined by \cite{Grossman:1996ke}
 \be
 \Delta_{\beta_d} = \sin2\beta_{J/\Psi K^0} -\sin2\beta_{\phi K^0} \label{eq:dbeta}
 \ed
 which is less than  $5\%$ in the SM \cite{Grossman:1996ke}. If a large value of $\Delta_{\beta_d}$ is measured, it will be a strong hint for new physics beyond SM.

\section{Numerical Analysis}\label{sec:NA}

So far, we have introduced seven new free parameters in the
general chiral color models and they are: two gauge couplings
$g_{V, A}$, four parameters in the two complex quantities
$F^{D}_{qb}$ and $m_V$.  In order to display the dependence of
$\Delta_{\beta_d}$ on $m_V$, we use the results in
Ref.~\cite{Frampton:2009rk} and take $g_V=-0.577g_s$ and
$g_A=-1.155g_s$ with $\alpha_s=g_s^2/4\pi =0.119$. Thus, the five
remaining parameters are $|F^D_{qb}|$, $\beta^{\rm NP}_q$ for q=d,
s and $m_V$.  We list the input values
used for numerical calculations in Table~\ref{tab:inputs}, where the relevant CKM matrix
elements $V_{tq}=\bar V_{tq}\exp(-i\beta_q)$ are obtained from the
UTfit Collaboration \cite{Bona:2009tn}, the decay constant of
$B_q$ is referred to the result given by HPQCD Collaboration \cite{Gamiz:2009ku}, the CDF and D$\O$ average value of $\Delta
m_{B_s}$ is from Ref.~\cite{TheHeavyFlavorAveragingGroup:2010qj} and the SM Wilson
coefficients of $b\to q q' \bar q'$  are obtained from
Ref.~\cite{BBL}. 
Other inputs are obtained from particle data group (PDG) \cite{Nakamura:2010zzi}.
\begin{table}[hptb]
\caption{Numerical inputs for the parameters in the SM.
 } \label{tab:inputs}
\begin{ruledtabular}
\begin{tabular}{cccccc}
  $\bar V_{td}$ & $\beta_d$ & $\bar V_{ts}$ & $\beta_s$ & $m_{B_d}$ & $m_{B_s}$
 \\ \hline
 $8.51(22)\times 10^{-3}$ & $(22\pm 0.8)^{\circ}$ & $ -4.07(22) \times 10^{-2}
 $ & $-(1.03\pm 0.06)^{\circ}$ & 5.28 GeV & 5.37 GeV \\ \hline\hline
 $f_{B_d} \sqrt{\hat B}_d$ [MeV]& $f_{B_s} \sqrt{\hat{B_s}}$ [MeV] & $f_{B_d}$ [MeV] & $f_{B_s}$ [MeV] & $S^{\rm exp}_{J/\Psi K^0}$& $\bar m_t(\bar m_t)$ \\ \hline
 $216\pm 15 $  & $266 \pm 18$  & $190\pm 13 $  & $231 \pm 15$  &
 $0.655 \pm 0.024$ & 163.8  GeV \\ \hline\hline
  $(\Delta m_{B_d})^{\rm exp}$ & $(\Delta m_{B_s})^{\rm exp}$ & $C_3$  & $C_4$ & $C_5$ & $C_6$  \\ \hline
$0.507 \pm 0.005$ ps$^{-1}$ & $17.78 \pm 0.12$ ps$^{-1}$  & $0.013$ & $-0.0335$ & 0.0095$$ & $-0.0399$  \\

\end{tabular}
\end{ruledtabular}
\end{table}

\begin{figure}[hptb]
\includegraphics*[width=5. in]{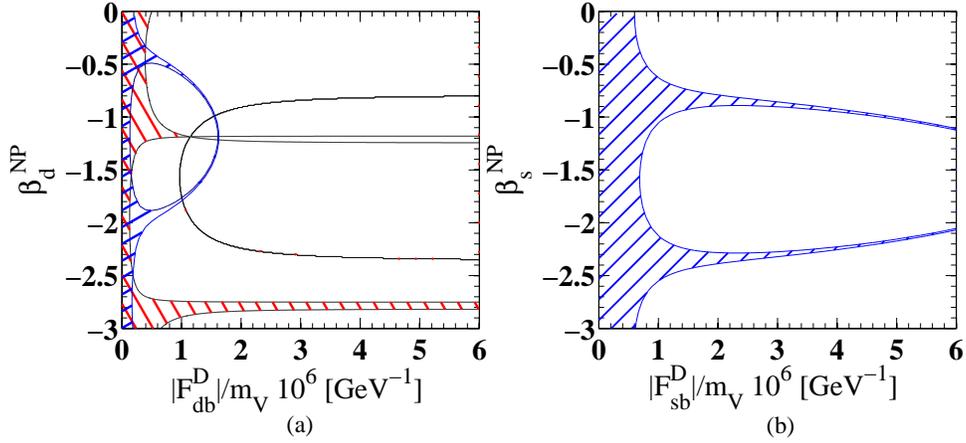}
\caption{(a)[(b)] Constraints on $\beta^{\rm NP}_{d[s]}$ and
$|F^D_{d[s]b}|/m_V$ (in units of $10^{-6}$) obtained from
$B_{d[s]}-\bar B_{d[s]}$ mixing (down-left hatched lines) and
$\sin2\beta_{J/\Psi K^0 }$ (down-right hatched lines). }
 \label{fig:constraints}
\end{figure}

\begin{figure}[bhpt]
\includegraphics*[width=4 in]{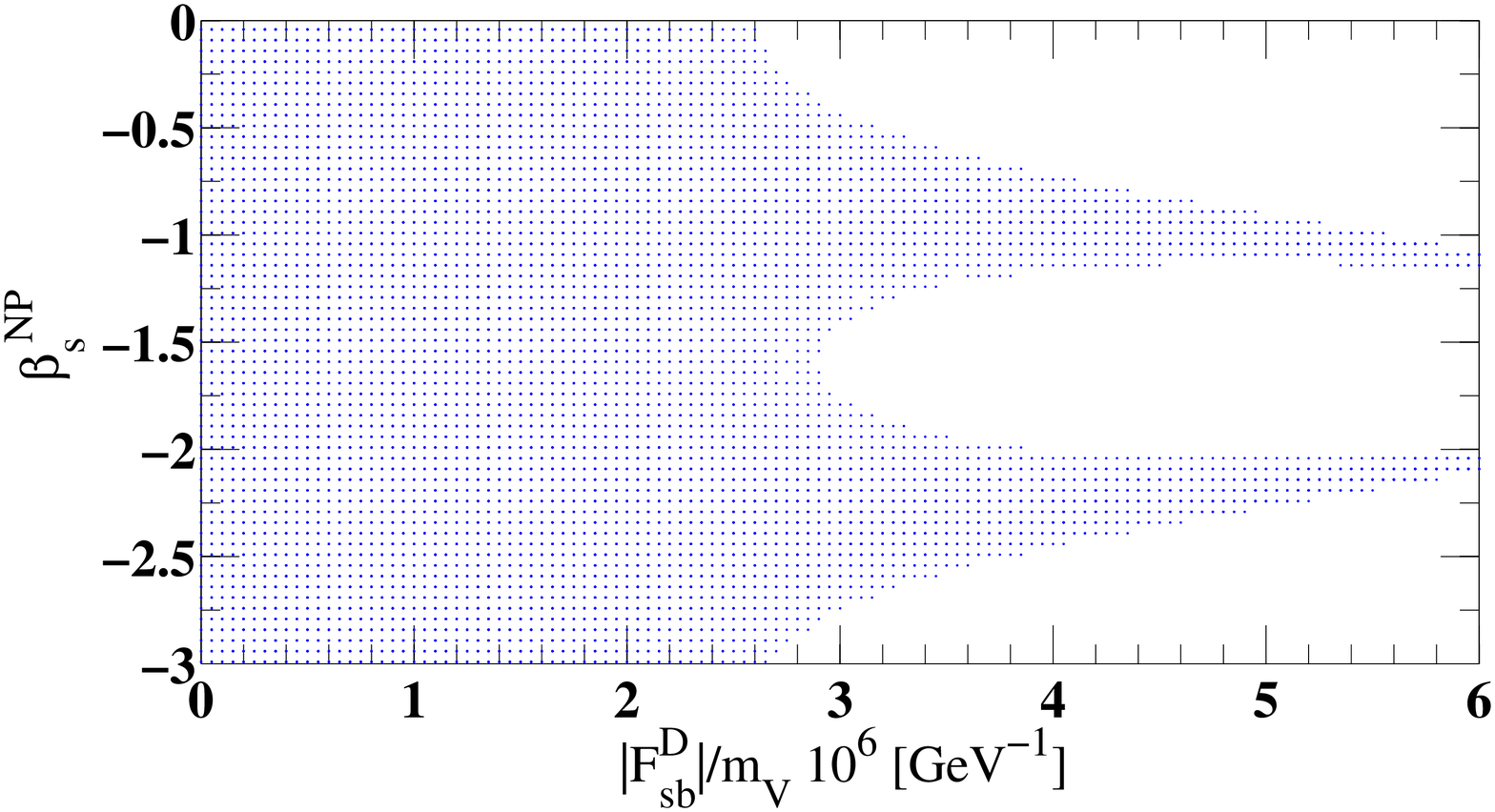}
\caption{Legend is the same as Fig.~\ref{fig:constraints}(b), but the errors of $\Delta m_{B_s}$ in the SM are included.}
 \label{fig:tot}
\end{figure}

After setting up the inputs, we study the contributions of the
axigluon to FCNC processes and their associated CPAs that are
defined earlier. We start by exploring the allowed parameter
space. Since the non-universal axigluon induces  FCNCs at  tree
level, the observed $B_q-\bar B_q$ mixing parameter $\Delta
m_{B_q}$ will give a strict constraint on the parameter space. In
Fig.~\ref{fig:constraints}(a)[(b)], the allowed range for
$\beta^{\rm NP}_{d[s]}$ and $|F^{D}_{d(s) b}|/m_V$ (in units of
$10^{-6}$) is drawn by the down-left hatched lines where we have
taken the SM contributions ($\Delta m^{\rm SM}_{B_d}$, $\Delta
m^{\rm SM}_{B_s}$) to be $(0.506,\, 17.76)$ ps$^{-1}$.
Furthermore, since the observed $S_{J/\Psi K^0}$ has been a
precise measurement, it is plausible that the current data can
further exclude the values of the parameter space which are
allowed by $\Delta m_{B_d}$. Taking $2\sigma$ errors of $S^{\rm
exp}_{J/\Psi K^0 }$ as the experimental bound, the allowed region
for $\beta^{\rm NP}_{d}$ and $|F^{D}_{db}|/m_V$ sketched by
down-right hatched lines  is plotted in
Fig.~\ref{fig:constraints}(a). Clearly, $S^{\rm exp}_{J/\Psi K^0 }$ gives a strong constraint on the parameters that contribute to $M^{B_d}_{12}$.  From Fig.~\ref{fig:constraints}, we see that,
except the two  narrow regions correspond to $|F^D_{db}|/m_V> 1\times 10^{-6}$ GeV$^{-1}$, the allowed values of $|F^D_{db}|/m_V$ are limited to be $|F^D_{db}|/m_V \leq 0.4\times 10^{-6}$ GeV$^{-1}$, whereas the allowed values of $|F^D_{sb}|/m_V$ can  be one order of magnitude larger than those of $|F^D_{db}|/m_V$. In general, the range of the CP violating phase $\beta^{\rm NP}_q$ is $[-\pi, \pi]$, for illustration, we just show the results within $[-\pi, 0]$. The pattern of the constraint in $[0,\pi]$ is similar to that in $[-\pi,0]$.  In order to illustrate the influence of the uncertainties of the SM on the free parameters, in Fig.~\ref{fig:tot} we plot the allowed values of $|F^D_{sb}|/m_V$ and $\beta^{\rm NP}_s$ by including the errors of $f_{B_s} \sqrt{\hat B_s}$ and $V_{ts}$. Comparing with Fig.~\ref{fig:constraints}(b), we see that the allowed range is extended slightly. We note that due to the strict constraint of $S^{\rm exp}_{J/\Psi K^0 }$, the bounds on the parameters for $b\to d$ transition are not changed significantly, therefore, we don't show the corresponding diagram for $b\to d$ transition. 

\begin{figure}[bhpt]
\includegraphics*[width=5.5 in]{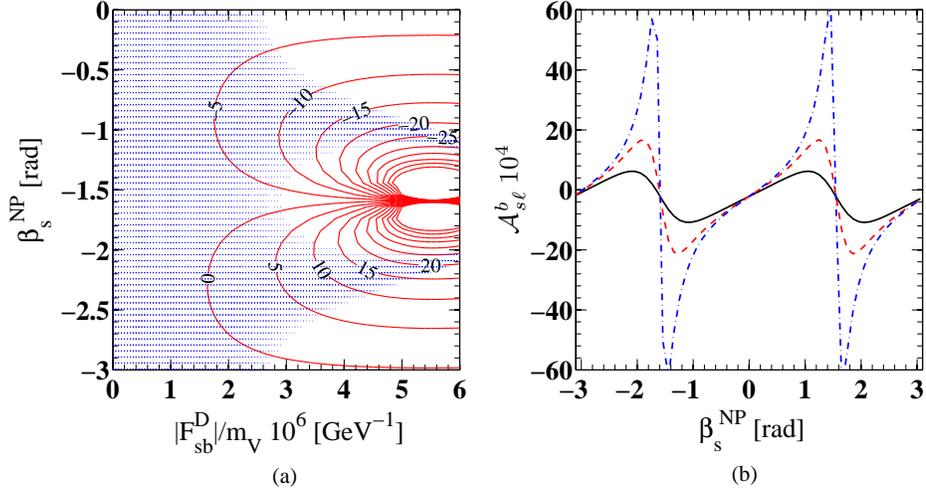}
\caption{(a) Contours of ${\cal A}^b_{s\ell}$ as a function of
$\beta^{\rm NP}_{s}$ and $|F^D_{sb}|/m_V$ (in units of $10^{-6}$).
(b) ${\cal A}^b_{s\ell}$ as a function of $\beta^{\rm NP}_s$,
where the solid, dashed and dash-dotted line stands for
$|F^D_{sb}|/m_V=(3, 4, 5)\times 10^{-6}$, respectively. The values
on the plot (a) are ${\cal A}^{b}_{s\ell}$ in units of $10^{-4}$.
}
 \label{fig:Absl}
\end{figure}

According to Eq.~(\ref{eq:Absl_1}), if we assume no new CP
violating phase in semi-leptonic decays, we will see that the
charge asymmetry ${\cal A}^b_{s\ell}$ depends on two kinds of CP
violating phases.  One of the two phases is originated  from
$B_d-\bar B_d$ mixing which is a $b\to d$ transition, and the
other phase is originated from $B_s-\bar B_s$ which is associated with  $b\to s $
transition. In other words, we have to consider four parameters
$\beta^{\rm NP}_{(d,s)}$ and $|F^{D}_{(d,s)b}|/m_V$
simultaneously.  However, if we consider $b\to (d, s)$ transitions
at the same time, we may induce a large effect on $s\to d$ because
the $\Delta K=2$ process is associated with $F^D_{ds}=(\zeta-1)
V^D_{13} V^{D*}_{23}$, i.e. $B_d-\bar B_d$, $B_s-\bar B_s$ and
$K^0-\bar K^0$ mixings have  strong correlations. In order to
avoid inducing  a large $K^0-\bar K^0$ mixing, we set a small
value for $V^D_{13}$.  This is consistent with the results
 shown in Fig.~\ref{fig:constraints}(a) where $\Delta m_{B_d}$ and $S_{J/\Psi
K^0 }$ strongly constrain  $|F^{D}_{db}|/m_V$. Hence, we assume
that $a^{d}_{s\ell}$ is dominated by the SM contribution where
$a^{d}_{s\ell}(SM)=-4.8\times 10^{-4}$ \cite{Lenz:2006hd}.
Consequently, the enhanced $|{\cal A}^{b}_{s\ell}|$ can be
attributed to $b\to s $ transition. With Eqs.~(\ref{eq:mBq}),
(\ref{eq:aqsl}) and (\ref{eq:Absl_2}) and the values given in
Table~\ref{tab:inputs}, the contours of ${\cal A}^{b}_{s\ell}$ as
a function of $\beta^{\rm NP}_{s}$ and $|F^{D}_{sb}|/m_V$ are
shown in Fig.~\ref{fig:Absl}(a) where the values of the contours
are in units of $10^{-4}$. As can be seen from the figure, not
only the sign of ${\cal A}^{b}_{s\ell}$ can fit the data, but also
its magnitude can be enhanced by axigluon-mediated effects. By
combining  with the constraint of $\Delta m_{B_s}$, the region of
$\beta^{\rm NP}_{s}$ for large $|{\cal A}^{b}_{s\ell}|$ is
limited. In Fig.~\ref{fig:Absl}(b), we display ${\cal
A}^{b}_{s\ell}$ as a function of $\beta^{\rm NP}_{s}$ where the
solid, dashed and dash-dotted line represents
$|F^{D}_{sb}|/m_V=(3, 4, 5)\times 10^{-6}$ GeV$^{-1}$,
respectively. As shown in the figure, negative and positive values
of $\beta^{\rm NP}_{s}$ can  enhance ${\cal A}^{b}_{s\ell}$. It should be noted that, although the axigluon-mediated effect can not enhance the like-sign charge asymmetry to be as large as the central value of D{\O} data, however, $|{\cal A}^{b}_{s\ell}|$ is
enhanced by one order of magnitude larger than the SM prediction.

\begin{figure}[hpbt]
\includegraphics*[width=5.5 in]{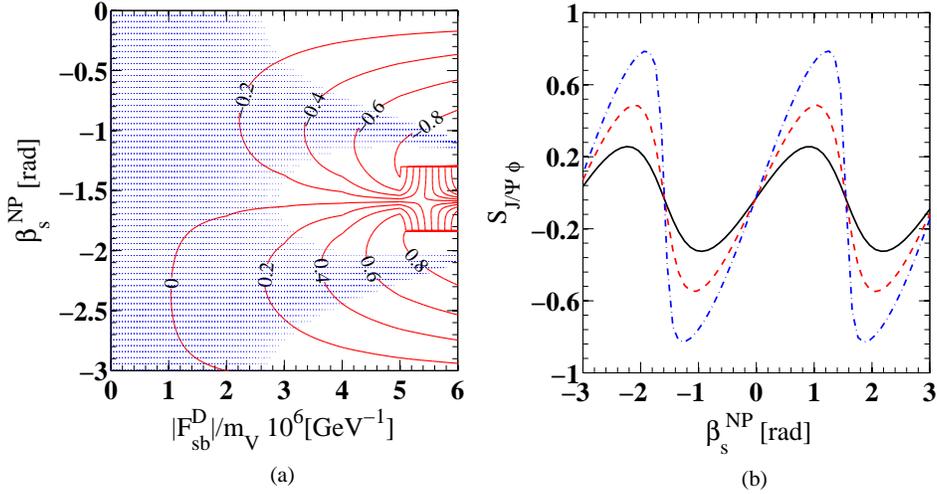}
\caption{ (a) Contours of $S_{J/\Psi \phi}$  as a function of
$\beta^{\rm NP}_s$ and $|F^D_{sb}|/m_V$ (in units of $10^{-6}$).
(b) $S_{J/\Psi \phi}$ as a function of $\beta^{\rm NP}_s$, where
the solid, dashed and dash-dotted line represents
$|F^D_{sb}|/m_V=(3,4,5)\times 10^{-6}$, respectively. }
 \label{fig:sjpsi_phi}
\end{figure}

Unlike the case of the charge asymmetry, the time-dependent CPA of
$B_s\to J/\Psi \phi$ decay depends only on the CP phase in $b\to
s$ transition. As a consequence, when the new CP violating effects
are small in $M^{B_d}_{12}$, ${\cal A}^{b}_{s\ell}$ and $S_{J/\Psi
\phi}$ defined in Eq.~(\ref{eq:Sf}) can have a strong correlation.
By using Eq.~(\ref{eq:Sjpsi_phi}), the contours of $S_{J/\Psi \phi}$ are plotted as a
function of $\beta^{\rm NP}_{s}$ and $|F^{D}_{sb}|/m_V$ in
Fig.~\ref{fig:sjpsi_phi}(a). From the figure, we find that
large $S_{J/\Psi \phi}$ can be archived  when ${\cal A}^{b}_{s\ell}$
is one order of magnitude larger than the SM prediction. Moreover,
we also plot $S_{J/\Psi \phi}$ as a function of $\beta^{\rm
NP}_s$ in Fig.~\ref{fig:sjpsi_phi}(b), where the solid, dashed and
dash-dotted line denotes $|F^D_{sb}|/m_V=(3, 4, 5)\times 10^{-6}$
GeV$^{-1}$, respectively. Clearly, a large ${\cal A}^{b}_{s\ell}$
indicates a large $S_{J/\Psi \phi}$. Although
 the measured values of ${\cal A}^{b}_{s\ell}$ and
 $S_{J/\Psi\phi}$ contain large errors, however, a few sigma deviations  from the SM
prediction can be considered as a hint for new physics effect.

\begin{figure}[tbhp]
\includegraphics*[width=5.5 in]{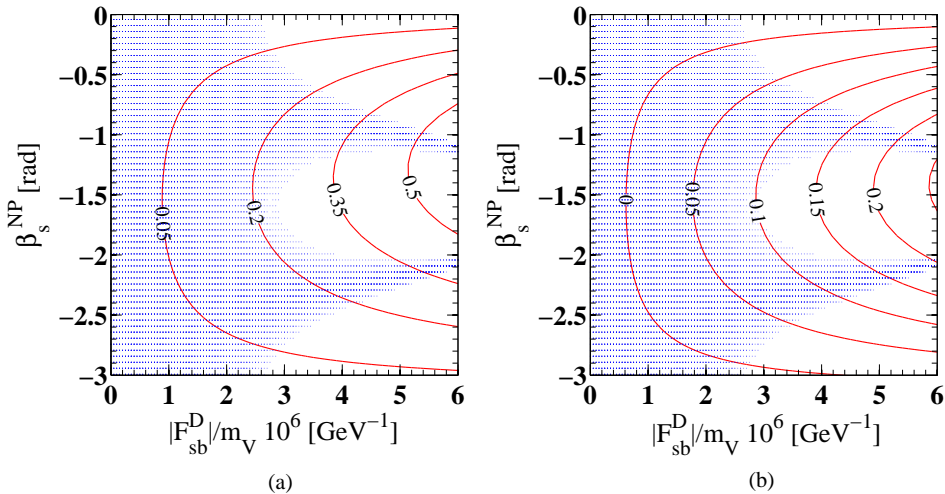}
\caption{ Contours of $\Delta_{\beta_d}$  as a function of
$\beta^{\rm NP}_s$ and $|F^D_{sb}|/m_V$ (in units of $10^{-6}$)
with (a) $m_V=0.5$ TeV and (b) $m_V=1$ TeV.}
 \label{fig:delta_jpsi_phi}
\end{figure}

\begin{figure}[tbhp]
\includegraphics*[width=5.5 in]{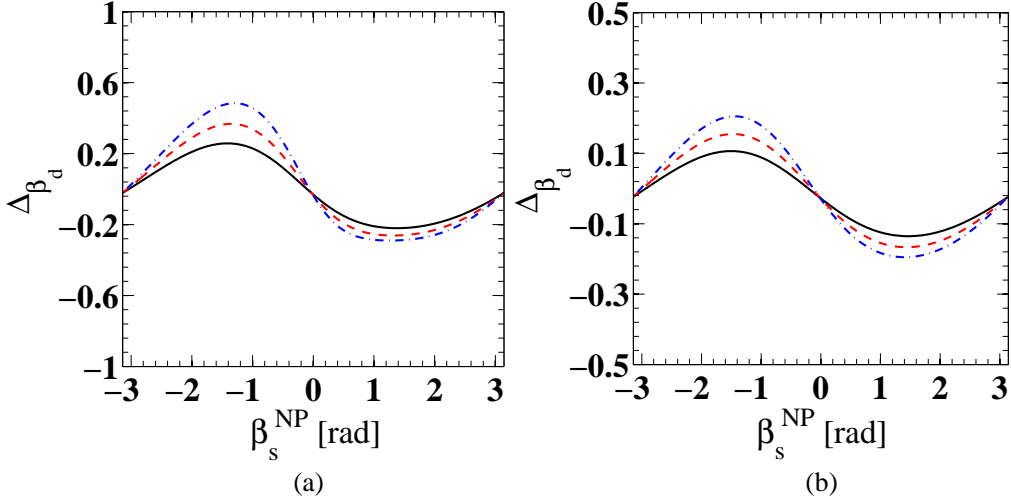}
\caption{ $\Delta_{\beta_d}$  as a function of $\beta^{\rm NP}_s$ with (a) $m_V=0.5$ TeV
and (b) $m_V=1$ TeV, where the solid, dashed and dash-dotted line represents $|F^D_{sb}|/m_V=(3,4,5)\times 10^{-6}$, respectively.  }
 \label{fig:delta_jpsi_phi_sin}
\end{figure}

It is well known that the golden process to measure the angle
$\beta_d$ in the SM is $B_d \to J/\Psi K^0 $ which  is dominated by
tree diagram. Although new physics can also affect this decay mode
via $b\to s c\bar c$ transition, however, as discussed in
Eq.~(\ref{eq:zeroCs}), the axigluon contributions to $B_d \to
J/\Psi K^0$ vanish. Hence, the source of the time-dependent CPA in
$B_d\to J/\Psi K^0$ decay is only originated from the $B_d$
oscillation. Since $\beta_d$ is also a parameter in the SM, a
single measurement of $S_{J/\Psi K^0}$ or $\sin 2\beta_{J/\Psi
K^0 }$ is hard to uncover the new physics. To probe the new physics, the
best way is to compare the CPA of $J/\Psi K^0$ with that of $\phi
K_S$. Therefore, we do not discuss each of $S_{J/\Psi K^0}$ and
$S_{\phi K^0}$ separately. Instead, we focus on the CPA difference
$\Delta_{\beta_d}$ which is defined in Eq.~(\ref{eq:dbeta}) and it
is only few percent in the SM. By Eqs.~(\ref{eq:S_phik}) and
(\ref{eq:dbeta}), we  see that although $\Delta_{\beta_d}$ is
insensitive to $F^{D}_{db}$ however it is strongly dependent  on
$F^D_{sb}$. To see the contributions of the axigluon to
$\Delta_{\beta_d}$, we present the contours of $\Delta_{\beta_d}$
as a function of $\beta^{\rm NP}_{s}$ and $|F^{D}_{sb}|/m_V$ in
Fig.~\ref{fig:delta_jpsi_phi}(a)[(b)], where we have set
$|F^{D}_{db}|/m_V=0$ and figure (a)[(b)]  corresponds to
$m_V=0.5[1]$ TeV.  Since the decay amplitude of $B\to \phi K$
depends on $F^D_{sb}/m^2_V$ while $\Delta m_{B_s}$ is
$(F^D_{sb}/m_V)^2$, thus a specific value for $m_V$ has to be
given when calculating the contours of $\Delta_{\beta_d}$. For
further understanding the $\beta^{\rm NP}_s$-dependence, we
display $\Delta_{\beta_d}$ as a function of $\beta^{\rm NP}_{s}$
in Fig.~\ref{fig:delta_jpsi_phi_sin}, where figure (a)[(b)] is for
$m_V=0.5[1]$ TeV and the solid, dashed and dash-dotted line stands
for $|F^{D}_{sb}|/m_V=(3,4,5)\times 10^{-6}$ GeV$^{-1}$,
respectively. It is clear that the axigluon contributions to
$\Delta_{\beta_d}$ are larger than that of the SM.
%

\begin{figure}[tbhp]
\includegraphics*[width=6 in]{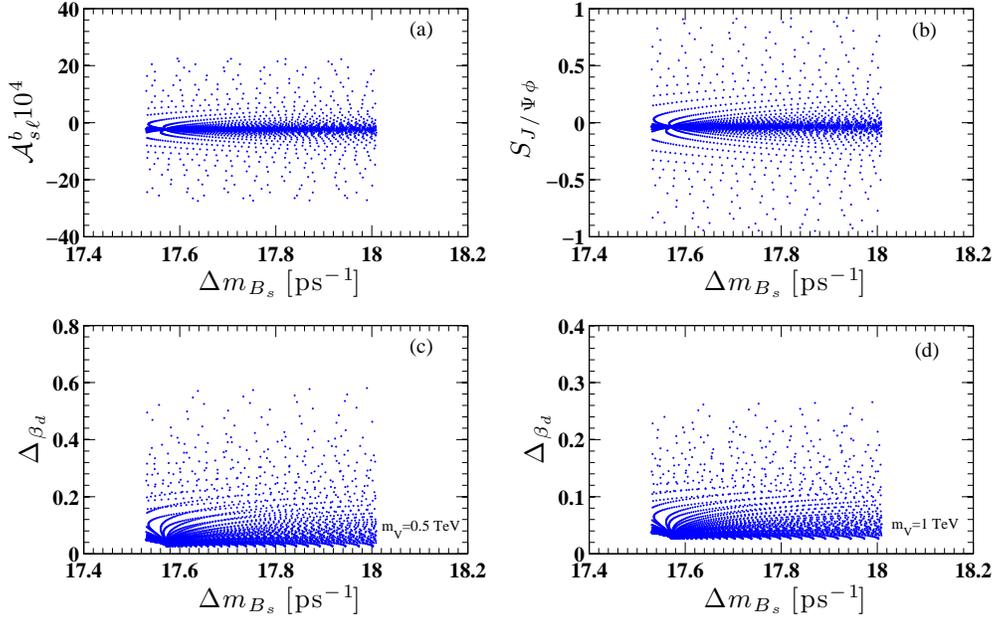}
\caption{ Correlations between $\Delta m_{B_s}$ and (a) ${\cal
A}^b_{s\ell}$, (b) $S_{J/\Psi \phi}$, (c)[(d)] $\Delta_{\beta_d}$
with $m_V=0.5 [1]$ TeV, where the angle $\beta^{\rm NP}_s$ belongs
to $[-\pi, 0]$.}
 \label{fig:correl1}
\end{figure}

\begin{figure}[tbhp]
\includegraphics*[width=5. in]{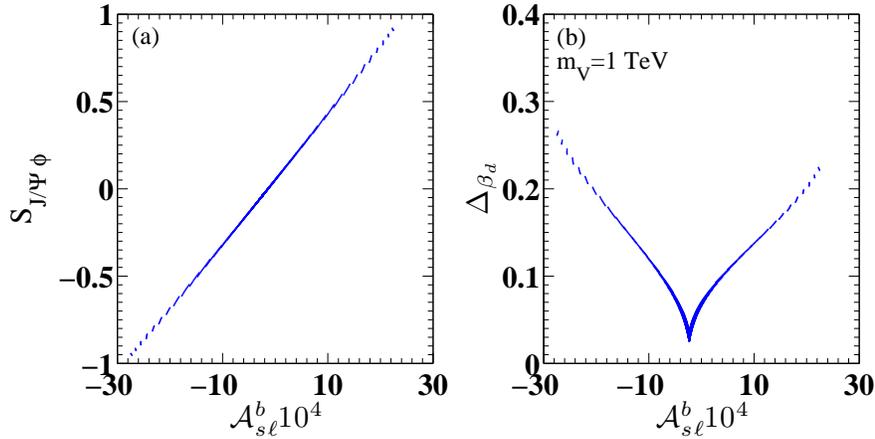}
\caption{ (a) Correlation between ${\cal A}^b_{s\ell}$ and
$S_{J/\Psi \phi}$ and (b) correlation between ${\cal A}^b_{s\ell}$
and $\Delta_{\beta_d}$ with $m_V=1$ TeV, where the constraint of
$\Delta m_{B_s}$ has been included and the angle $\beta^{\rm
NP}_s$ belongs to $[-\pi, 0]$.}
 \label{fig:correl2}
\end{figure}

 In order to  comprehend further the correlations
among various physical observables under the influence of the
axigluon, we display the scatter plots of ${\cal A}^b_{s\ell}$,
$S_{J/\Psi \phi}$ and  $\Delta_{\beta_d}$ with $m_V=0.5 (1)$ TeV
versus  $\Delta m_{B_s}$ in Fig.~\ref{fig:correl1}, where we have
chosen the range of $\beta^{\rm NP}_{s}$ to be $[-\pi, 0]$. As an
illustration, we also show the scatter plots of ( ${\cal
A}^b_{s\ell}$, $S_{J/\Psi \phi}$) and (${\cal A}^b_{s\ell}$,
$\Delta_{\beta_d}$) with $m_V=1$ TeV in Fig.~\ref{fig:correl2}, in
which the constraint of $\Delta m_{B_s}$ has been included and
$\beta^{\rm NP}_{s}$ belongs to $[-\pi, 0]$. By
Fig.~\ref{fig:correl2}(a), we see that the correlation between
${\cal A}^b_{s\ell}$ and $S_{J/\Psi \phi}$ is linear, where this
behavior can be understood by the linear dependence between  the
like-sign charge asymmetry and the mixing-induced CPA of $B_s$.
Due to the linearity, we expect that the correlation between
$S_{J/\Psi \phi}$ and $\Delta_{\beta_d}$ should be similar to that
between ${\cal A}^b_{s\ell}$ and $\Delta_{\beta_d}$. Therefore, we
just show the latter case in Fig.~\ref{fig:correl2}(b).

\section{Conclusion}\label{sec:conclusion}

In general, a flavor non-universal axigluon in generalized chiral
color models  can  induce FCNCs at tree level. We study
phenomenologically  the axigluon-mediated effects on $\Delta B=2$
FCNC processes and the associated CPAs. We find that although
$\Delta m_{B_q}$  strongly constrain the free parameters, the
precise measurement of $S_{J/\Psi K^0}$  can further exclude the
parameter space of $b\to d$ transition.  Furthermore, for avoiding
inducing large $K^0-\bar K^0$ mixing,  the parameter $V^D_{13}$ is
chosen to be small so that the like-sign charge asymmetry ${\cal
A}^{b}_{s\ell}$ and $\Delta_{\beta_d}$ are insensitive to the
parameters of $b\to d$ transition.  As a result, the CP violating
observables ${\cal A}^{b}_{s\ell}$, $S_{J\Psi \phi}$ and
$\Delta_{\beta_d}$ are strongly correlated and are only sensitive
to the parameters of $b\to s$ transition.

By the study, we find that the axigluon effects do not only
preserve the negative sign in ${\cal A}^{b}_{s\ell}$, but also
enhance its magnitude by one order of magnitude larger than the SM
prediction. Subsequently, the associated values of the parameters
can also enhance the CPA $S_{J/\Psi \phi}$ and the CPA difference
$\Delta_{\beta_q}$ largely although they are only few percent in
the SM.

\section*{Acknowledgement}

This work is supported by the National Science Council of R.O.C.
under Grant No. NSC-97-2112-M-006-001-MY3. The author C.H.C would
like to thank Prof. Young-Chung Hsue for his help on using plot
tool. G. Faisel would like thank the National Center for
Theoretical Sciences (NCTS) at Cheng Kung University for the
hospitality where this work has been done.

\end{document}